\documentclass[12pt]{article}
\let\section=\subsection     \let\subsection=\subsubsection                
\usepackage{graphicx}

\begin{document}
\begin{center}
   {\large \bf From meson- and photon-nucleon scattering to vector}\\[2mm]
   {\large \bf mesons in nuclear matter}\\[5mm]
   M.F.M. Lutz, \underline{Gy. Wolf}\footnote{permanent address:RMKI KFKI, Pf. 49, H-1525 Budapest, Hungary}, B.~Friman \\[5mm]
   {\small \it  Gesellschaft f\"ur Schwerionenforschung (GSI) \\
   Postfach 110552, D-64220 Darmstadt, Germany \\[8mm] }
\end{center}

\begin{abstract}\noindent
We present a relativistic and unitary approach to
pion- and photon-nucleon scattering  taking into account the $\pi N$,
$\rho N$, $\omega N$, $\eta N$, $\pi\Delta$, $K \Lambda$ and $K \Sigma$ 
channels.
Our scheme dynamically generates the s- and d-wave nucleon resonances
$N(1535)$, $N(1650)$, $N(1520)$ and  $N(1700)$ and isobar
resonances $\Delta(1620)$ and  $\Delta(1700)$ in terms of quasi-local two-body
interaction terms. We obtain a fair description of the experimental data
relevant for slow vector-meson propagation in nuclear matter. The
s-wave $\rho $- and $\omega $-meson nucleon scattering amplitudes,
which define the leading density modification of the $\rho$- and 
$\omega $-meson spectral functions in nuclear matter, are predicted.
\end{abstract}

\section{Introduction}

The in-medium properties of hadrons is a topic of high current
interest. The decay of vector mesons into $e^+e^-$ and $\mu^+\mu^-$ pairs
offers a unique tool to explore the properties of dense and hot
matter in nuclear collisions. The lepton pairs provide virtually
undistorted information on the current-current correlation function
$\langle j_\mu j_\nu\rangle$ in the medium~\cite{RW}. At invariant
masses in the range $500 - 800$ MeV, $\langle j_\mu j_\nu\rangle$
is sensitive to in-medium modifications of the mass distribution of
the light vector mesons $\rho$ and $\omega$.
There is a longstanding and controversial discussion about the vector
meson properties in dense nuclear matter~\cite{su-huong,BR,HFN,Wambach}. 
To leading orders
in the baryon density, modifications of the mass distribution are
determined by the vector-meson nucleon scattering amplitudes. Since
these amplitudes are not directly constrained by data, the predictions
for the vector-meson spectral densities in nuclear matter are strongly
model dependent.

This work is an attempt to overcome this problem by using the
constraints from data on pion- and photon-nucleon scattering
considering in particular the $\omega$- and $\rho$-meson production data
in a systematic way. We construct a coupled-channel scheme for
meson-baryon scattering, including the $\pi N$, $\rho N$, $\omega
N$, $\pi \Delta$, $\eta N$, $K\Lambda$ and $K\Sigma$ channels. In
such a scheme the amplitudes for experimentally non-accessible
processes like $\rho N$ and $\omega N$ scattering are constrained
by the data on elastic $\pi N$ scattering and inelastic reactions like the
pion- and photon-induced production of vector mesons. Our goal is to determine
the vector-meson nucleon scattering amplitudes close to threshold. Consequently we
concentrate on the energy window 1.4 GeV $< \sqrt{s} <$ 1.8 GeV. It is
sufficient to consider only s-wave scattering in the $\rho
N$ and $\omega N$ channels. This implies that in the $\pi N$ and
$\pi\Delta$ channels we need only s- and d-waves. In particular, we
consider the $S_{11}, S_{31}, D_{13}$ and $D_{33}$ partial waves of
$\pi N$ scattering. In order to systematically derive the momentum
dependence of the vector-meson self energy, vector-meson nucleon
scattering also in higher partial waves would have to be
considered.

\section{Relativistic coupled-channel dynamics}

Here we only outline our model, the detailed discussion can
be found in \cite{LWF}.
We construct an effective Lagrangian with quasi-local four-point
meson-baryon contact interactions in order to
study the meson-baryon scattering process. Within this
framework, the Bethe-Salpeter equation for the coupled-channel
system reduces to a matrix equation.

The general four-point interaction Lagrangian for the coupled-channel problem
in momentum space can be written in the form
\begin{eqnarray}
&&{\mathcal L}(\bar k ,k ;w)=
\sum_{I}\,R^{(I)\,\dagger }(\bar q,\bar p)\,\gamma_0
\,K^{(I)}(\bar k ,k ;w )\,R^{(I)}(q,p)\,+ \dots\,,\;\;\;\;
\label{eff-lagr}
\end{eqnarray}
where $K^{(I)}_{ab}(\bar k , k;w)$ is a channel dependent interaction matrix,
which in general is a function of the relative momenta of
the initial and final states $k=\frac 12 (p-q)$ and $\bar k=\frac
12 (\bar p-\bar q)$ as well as the total momentum $w = p+q = \bar p
+ \bar q$. The state vectors $R^{(I)}(q,p)$ involve the positive frequency parts
of the fields specifying the various channels.

The coupled-channel Bethe-Salpeter equation projected onto a given
isospin then reads
\begin{eqnarray}
T^{(I)}_{ab}(\bar k ,k ;w ) &=& K^{(I)}_{ab}(\bar k ,k ;w )
+\sum_{c,d}\int\!\! \frac{d^4l}{(2\pi)^4}\,K^{(I)}_{ac}(\bar k , l;w )\,
G^{(I)}_{cd}(l;w)\,T^{(I)}_{db}(l,k;w )\;,
\nonumber\\
G^{(I)}_{cd}(l;w)&=&-i\,D_{\Phi(I,c)\Phi(I,d)}(\frac{1}{2}\,w-l)\,S_{B(I,c)B(I,d)}(
\frac{1}{2}\,w+l)\,,
\label{BS-coupled}
\end{eqnarray}
where $D_{\Phi(I,c)\Phi(I,d)}(q)$ and $S_{B(I,c)B(I,d)}(p)$ denote
the meson and baryon Feynman propagators in a given channel of isospin $I$.
A convenient labelling of the channels is obtained by defining
\begin{eqnarray}
&&\Phi({\textstyle{1\over 2}},a)=(\pi,\pi,\rho_\mu ,\omega_\mu,\eta ,K, K)_a\;,\;\;\;\;\;\;\
B({\textstyle{1\over 2}},a)=(N,\Delta_\mu,N, N,N,\Lambda , \Sigma )_a\;,
\nonumber\\
&&\Phi({\textstyle{3\over 2}},a)=(\pi,\pi, \rho_\mu, K)_a\;,\;\;\qquad  \quad \quad  \;\,\;\;
B({\textstyle{3\over 2}},a)=(N, \Delta_\mu,N,\Sigma )_a\;.
\label{def-channel}
\end{eqnarray}

The general form of the interaction
kernel may be obtained from a meson exchange
model or more systematically from the chiral Lagrangian. In this paper we
construct an effective field theory for meson-nucleon scattering in
the resonance region. The philosophy of this approach is to
approximate the interaction kernel $K$ but to treat rescattering in the
s-channel explicitly. In our scheme, we assume that the interaction
kernel is slowly varying in energy in the relevant window 1.4 GeV$
<\sqrt{s}< 1.8$ GeV.

We constrain our model parameters by data from photon-induced meson
production off the nucleon.
According to the original conjecture of Sakurai \cite{Sakurai} the vector-meson converts into a real photon
via the interaction terms $A^\mu\,\rho_\mu^{(0)} $ and $A^\mu\,\omega_\mu$. In our scheme these terms
are not allowed. In order to simulate the successful
vector-meson dominance assumption it is therefore natural to directly relate 
the strength of the
$\gamma N \to X $ and $\rho^{(0)}\,N \to X$, $\omega\,N \to X$ vertices, where $X$ is any hadronic final state.

We specify the generalized form of the vector-meson dominance conjecture as 
applied for the direct photon-induced production vertices 
$K^{(I),\mu}_{\gamma N \rightarrow X}(\bar q,q;w)$:
\begin{eqnarray}
&& \frac{1}{2} \left(
K^{(\frac{1}{2}),\,\mu}_{\gamma \,p \rightarrow X} +
K^{(\frac{1}{2}),\,\mu}_{\gamma \,n \rightarrow X} \right) (\bar q,q;w)=
e \, K^{(\frac{1}{2})}_{\nu, \,\omega N \rightarrow X}(\bar q,q;w)
\,\Gamma^{\nu \mu}_S (q; w)\,,
\nonumber\\
&& \frac{1}{2} \left(
K^{(\frac{1}{2}),\,\mu}_{\gamma \,p \rightarrow X} -
K^{(\frac{1}{2}),\,\mu}_{\gamma \,n \rightarrow X} \right) (\bar q,q;w)=
\frac{e}{\sqrt{3}} \, K^{(\frac{1}{2})}_{\nu, \,\rho N \rightarrow X}(\bar q,q;w)
\,\Gamma^{\nu \mu}_V (q; w)\,,
\nonumber\\
&& K^{(\frac{3}{2}),\,\mu}_{\gamma \,p \rightarrow X} (\bar q,q;w)=
e \sqrt{{\textstyle{2\over 3}}}\, \,
K^{(\frac{3}{2})}_{\nu,\, \rho N\rightarrow X}(\bar q,q;w)\,\Gamma^{\nu \mu}_V (q;w)\,,
\nonumber\\
&& K^{(\frac{3}{2}),\,\mu}_{\gamma \,n \rightarrow X} (\bar q,q;w)=
e \sqrt{{\textstyle{2\over 3}}}\, \,
K^{(\frac{3}{2})}_{\nu,\, \rho N\rightarrow X}(\bar q,q;w)\,\Gamma^{\nu \mu}_V (q;w)\,,
\label{gamma-ansatz:k}
\end{eqnarray}
where $X$ stands for any hadronic two-body final state with isospin $I$.
The transverse objects $\Gamma_{S(V)}^{\mu \nu}(q; w)$ are the most general
transition tensors compatible with gauge invariance. 
Its form can be found in \cite{LWF}.

\section{Results}

In accordance with the effective field theory approach only data in an
appropriate kinematical window is used in the analysis. The threshold for
vector-meson production off a nucleon is at $\sqrt{s} \simeq 1.7$
GeV. Close to the elastic pion-nucleon threshold we do not expect
our scheme to be efficient.  We therefore fit the data in the energy range
$1.4$ GeV $\leq \sqrt{s} \leq 1.8$ GeV, with the energy independent coupling
matrices.
Our model includes four coupled-channel matrices  carrying quantum 
numbers $I=\frac{1}{2}\,,\frac{3}{2}$ and $J=\frac{1}{2}\,,\frac{3}{2}$.

The set of parameters is adjusted to describe the partial-wave pion-nucleon 
phase shifts including their inelasticity parameters. Furthermore the pion- 
and photon-induced production cross sections of pseudo-scalar meson  and 
vector mesons are considered.
Of crucial importance to our fit are the empirical pion- and photon-induced
vector-meson production data. Only via the use of those data it is 
ultimately possible to determine the vector-meson nucleon scattering 
amplitudes, the prime goal of this work. All together we include $N=1156$ data
points. We obtained a fair description of the experimental data \cite{LWF}.

As a result of our fit the s- and d-wave baryon resonances 
$N(1535)$, $N(1650)$, $N(1520)$, $N(1700)$,
$\Delta(1620)$ and $\Delta(1700)$ are generated dynamically in our scheme.
In particular we
find that the s-wave resonances $N(1535)$ and $N(1650)$ show a strong
$\omega N$ component, which are required to simultaneously generate
both resonances. This is important because the two resonances mix strongly.
On the other hand, only the $N(1535)$ but not the $N(1650)$ resonance
appears to couple significantly to the $\rho N$ channel. Similarly striking
is our result that the d-wave $N(1520)$ resonance couples strongly to the
$\omega N$ channel but with only much reduced strength to the $\rho N$ channel.
This result is a consequence of our systematic inclusion of the photon-induced
scattering data not done previously.

\begin{center}
   \includegraphics[height=6.cm,angle=-00]{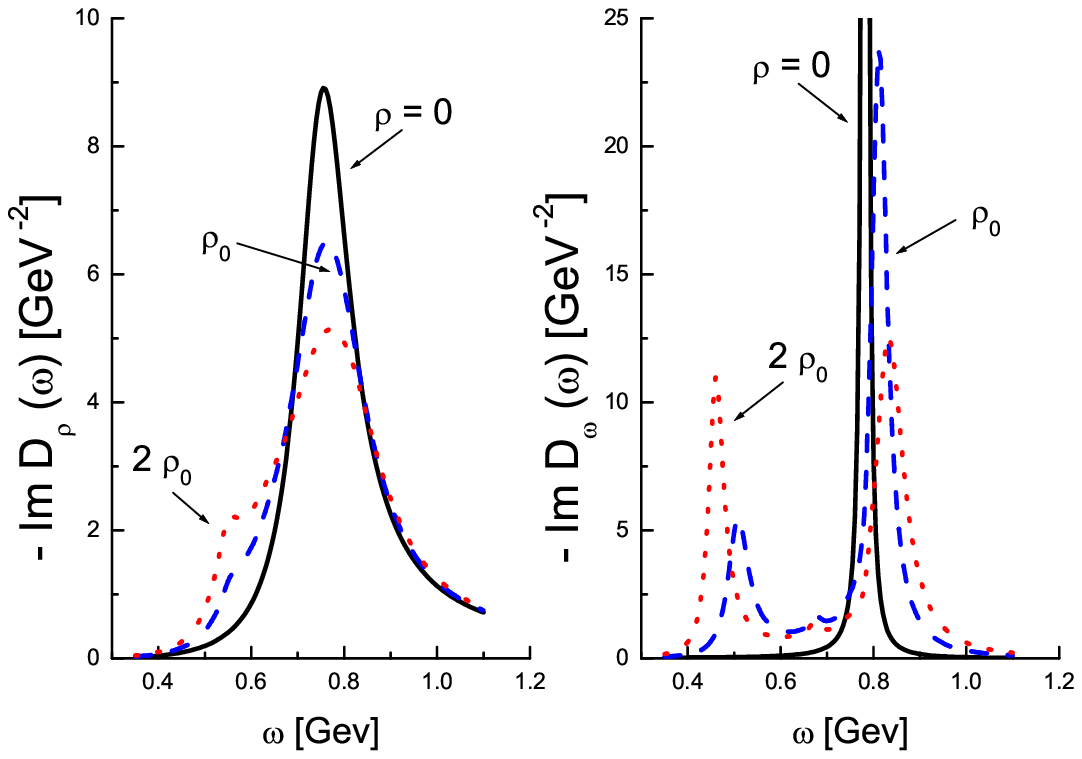}\\
  \parbox{14cm}
       {\centerline{\footnotesize 
       Fig.~1: Imaginary parts of the
$\rho$ and $\omega$ propagators in nuclear matter
}} \\ {\centerline{\footnotesize 
at $\rho=\rho_0$ and $2 \rho_0$, compared to those of the free-space propagators.}}
\label{propagators}
\end{center}
We apply the resulting vector-meson scattering amplitudes to the propagation 
of the $\rho$ and $\omega$ mesons in nuclear matter. 
Using the low-density theorem the self energy of a vector meson in nuclear 
matter is expressed in terms of the s-wave scattering amplitude averaged 
over spin and isospin \cite{LDT}.

In Fig.~\ref{propagators} we show the vector-meson propagators at the
saturation density of nuclear matter, $\rho_0 = 0.17
\mbox{~fm}^{-3}$ and at $\rho = 2 \rho_0$. For the $\rho$ meson
we note an enhancement of the width, and a downward shift in
energy, due to the mixing with the baryon resonances at $\sqrt{s} =
1.5 - 1.6$ GeV. At $\rho=\rho_0$ the
center-of-gravity of the spectral function is shifted down in
energy by about $3$~\%.
The in-medium propagator of the $\omega$ meson exhibits two
distinct quasiparticles, an $\omega$ like mode, which is shifted up
somewhat in energy, and a resonance-hole like mode at low energies.
The low-lying modes carry about 15 \% on the energy-weighted sum
rule. The center-of-gravity is shifted down by about $ 4$~\%.
However, we stress that the structure of the in-medium
$\omega$-meson spectral function clearly cannot be characterized by this
number alone.

A simple estimate on the accuracy of the leading order result follows by
investigating the size of the leading correction term determined by the Pauli blocking effect.
Take for example the $\omega$ meson for which its leading and subleading mass shift
terms require only the s-wave scattering length $a_{\omega   N}^{\frac{1}{2}}=(-0.45+i 0.31) $
fm and $a_{\omega N}^{\frac{3}{2}}=(-0.43+i0.15 )$ fm \cite{LWF}.
At nuclear saturation density the correction term
of order $k_F^4$  implies a further repulsive mass shift of $1$ MeV
and increase of the decay width of $2$ MeV for the $\omega $ meson.
We expect that the results obtained with only the leading term in
the low-density expansion are qualitatively correct at normal
nuclear matter density. However, on a quantitative level, the
spectral functions may change when higher order terms in the
density expansion are included. For instance, we expect that the
in-medium properties of the baryon resonances depend sensitively on
the meson spectral functions. If this is the case, a self
consistent calculation, which corresponds to a partial summation of
terms in the density expansion, would have to be
performed~\cite{Lutz:Korpa}.

\section{Summary and conclusion}

In this work we predicted the amplitudes which describe the s-wave scattering of the
light vector mesons off nucleons. To leading order in a density expansion these
amplitudes determine the spectral functions of the $\rho$- and $\omega$-meson
in nuclear matter. Since there exist no data on
vector-meson nucleon scattering we constrained our analysis by all relevant
elastic and inelastic $\gamma N$ and $\pi N$ data. The coupled channel
unitarity condition together with the causality property of local quantum field
theory then leads to quite robust predictions for the vector-meson nucleon scattering
amplitudes. Our amplitudes show rapid energy variations due to the presence of
nucleon and isobar resonances. In our novel and covariant coupled channel approach,
which considers the $\gamma N $, $\pi N$, $\pi \Delta $, $\rho N$,
$\omega N$, $\eta N$, $K \Lambda$ and $K \Sigma $ states, the s- and d-wave nucleon and
isobar resonances are generated dynamically by coupled channel effects.
In nuclear matter our scattering amplitudes lead to a $\omega$-meson spectral
functions with considerable support at energies smaller than the free-space mass
representing resonance nucleon-hole type excitations. On the other hand,
as an immediate consequence of the only moderate coupling of the $\rho N$ channel
to the $N(1520)$ d-wave resonance, the in-medium effects for the $\rho$ meson
are found to be significantly smaller as compared to previous works.

The results of our work are relevant for the experimental program at GSI. 
The HADES detector will help to further explore the properties of the light 
vector mesons in nuclear matter by measuring their dilepton final state with 
high accuracy. Complementary experimental programs are pursued at MAMI and KEK
with photon and nucleon induced reactions off nuclei. To further substantiate 
the structure of the vector-meson nucleon scattering amplitudes it would be 
desirable to establish a more microscopic understanding of the effective
interaction vertices employed in our work.

This work were partly supported by Hungarian Research Foundation (OTKA)
grants: T 26543 and T 30855.


\end{document}